\documentclass[aps,prx,twocolumn,superscriptaddress,longbibliography]{revtex4-2}

\usepackage{amssymb}
\usepackage{amsmath}
\usepackage{graphicx}
\usepackage[dvipsnames]{xcolor}
\usepackage{bm}
\usepackage{dcolumn}
\usepackage{dsfont}
\usepackage[hidelinks]{hyperref}
\usepackage{physics}
\usepackage{siunitx}

\usepackage{cleveref}
\crefname{equation}{Eq.}{Eqs.}
\Crefname{equation}{Equation}{Equations}
\crefname{figure}{Fig.}{Figs.}
\Crefname{figure}{Figure}{Figures}
\crefname{section}{Sec.}{Secs.}
\Crefname{section}{Section}{Sections}
\crefname{appendix}{Appendix}{Apps.}
\Crefname{appendix}{Appendix}{Apps.}
\crefname{paragraph}{Sec.}{Secs.}
\crefname{table}{Table}{Tables}

\usepackage{multirow}
\usepackage{tabularx}

\usepackage{hhline}

\begin{document}

\title{Quantum-tailored machine-learning characterization of a superconducting qubit}

\author{\'Elie Genois}
\email{elie.genois@usherbrooke.ca}
\address{Institut quantique \& D\'epartement de Physique, Universit\'e de Sherbrooke, Qu\'ebec J1K 2R1, Canada}
\author{Jonathan A. Gross}
\thanks{jarthurgross@google.com}
\address{Institut quantique \& D\'epartement de Physique, Universit\'e de Sherbrooke, Qu\'ebec J1K 2R1, Canada}
\author{Agustin Di Paolo}
\thanks{adipaolo@mit.edu}
\address{Institut quantique \& D\'epartement de Physique, Universit\'e de Sherbrooke, Qu\'ebec J1K 2R1, Canada}
\author{Noah J. Stevenson}
\address{Quantum Nanoelectronics Laboratory, University of California, Berkeley, Berkeley CA 94720, USA}
\author{Gerwin Koolstra}
\address{Quantum Nanoelectronics Laboratory, University of California, Berkeley, Berkeley CA 94720, USA}
\address{Computational Research Division, Lawrence Berkeley National Laboratory, Berkeley CA 94720, USA}
\author{Akel Hashim}
\address{Quantum Nanoelectronics Laboratory, University of California, Berkeley, Berkeley CA 94720, USA}
\author{Irfan Siddiqi}
\address{Quantum Nanoelectronics Laboratory, University of California, Berkeley, Berkeley CA 94720, USA}
\address{Computational Research Division, Lawrence Berkeley National Laboratory, Berkeley CA 94720, USA}
\address{Materials Science Division, Lawrence Berkeley National Laboratory, Berkeley CA 94720, USA}
\address{Canadian Institute for Advanced Research, Toronto, Ontario M5G 1M1, Canada}
\author{Alexandre Blais}
\address{Institut quantique \& D\'epartement de Physique, Universit\'e de Sherbrooke, Qu\'ebec J1K 2R1, Canada}
\address{Canadian Institute for Advanced Research, Toronto, Ontario M5G 1M1, Canada}

\begin{abstract}
    Machine learning (ML) is a promising approach for performing challenging quantum-information tasks such as device characterization, calibration and control.
    ML models can train directly on the data produced by a quantum device while remaining agnostic to the quantum nature of the learning task.
    However, these generic models lack physical interpretability and usually require large datasets in order to learn accurately.
    Here we incorporate features of quantum mechanics in the design of our ML approach to characterize the dynamics of a quantum device and learn device parameters.
    This physics-inspired approach outperforms physics-agnostic recurrent neural networks trained on numerically generated and experimental data obtained from continuous weak measurement of a driven superconducting transmon qubit.
    This demonstration shows how leveraging domain knowledge improves the accuracy and efficiency of this characterization task, thus laying the groundwork for more scalable characterization techniques.
\end{abstract}

\date{\today}
\maketitle

\section{Introduction}\label{sec:Intro}
Machine learning (ML) has recently been applied to solve problems in numerous areas of physics, including quantum-information science~\cite{carleo_machine_2019, Dunjko_MLinQuantum_2018}.
For example, ML models have been used to tackle quantum-computing tasks including qubit readout~\cite{lienhard_deep_2021, seif_machine_2018, ding_fast_2019}, quantum control~\cite{ostaszewski_approximation_2019, bukov_reinforcement_2018} and quantum state tomography~\cite{xu_neural_2018, torlai_neural-network_2018, Carrasquilla_MLquantummatter_2020} among others.
These ``black-box'' models enjoy wide applicability to different problems, as their 
structure is agnostic to the physical processes involved in the targeted task.
However, information about the physics of the learned process can also be leveraged in an effort to improve upon these black-box approaches by making them more trainable and interpretable.
Recent work has shown promising results in this direction
by using ``white-box'' quantum features, such as physical laws, symmetries and relevant correlations, in the ML approach.
For example, including quantum features has been used to improve the characterization of quantum noise~\cite{youssry_characterization_2020} as well as to learn quantum states more efficiently~\cite{Hibat-Allah_RNN_2020, morawetz_u1_2020} and in an interpretable way~\cite{valenti_correlation-enhanced_2021}.

Here we are interested in developing quantum-aware ML models in order to characterize the dynamics of a quantum device from physical observations.
Accurate device characterization is crucial for producing high-fidelity quantum operations and algorithms~\cite{preskill_quantum_2018}.
However, just like most techniques in quantum characterization, validation and verification, our ability to perform this characterization is intrinsically limited by quantum properties such as the lack of direct access to the state~\cite{eisert_quantum_2020, supic_self-testing_2020, carrasco_theoretical_2021}.
By tailoring the machine-learning approach to reflect as much of our knowledge of the quantum device as possible, we seek to mitigate the effect of these intrinsic difficulties.

As an example of a generic situation in quantum information, we focus on the circuit QED architecture and characterize the dynamics of a resonantly driven superconducting transmon qubit~\cite{Koch_transmon_2007}.
The driven qubit is continuously monitored by a weak dispersive measurement~\cite{blais_RMP_2020}.
By training a ML model to predict measurement outcomes from the weak-measurement data, we can characterize the qubit parameters via the quantum-trajectory formalism~\cite{wiseman_quantum_2010}.

This physics-inspired approach overcomes two important limitations of generic machine-learning tools that are agnostic to the physics of the problem at hand, such as Ref.~\cite{flurin_using_2020} which also considered learning quantum trajectories from the continuous weak measurement of a transmon qubit.
First, it is more efficiently trainable as significantly less experimental data is needed to train the ML model.
Second, the physics-inspired approach allows for a direct device characterization, whereas the black-box approach lacks interpretability as the trained ML model cannot be easily associated with an explicit physical description of the device. 
We address these two limitations by designing ML models that exploit our domain knowledge of quantum trajectories and dispersive measurements in circuit QED~\cite{blais_RMP_2020}.
We also explore the relations between the training efficiency, physical interpretability and achievable accuracy of the ML model predictions.

In the following, we train our ML models on numerically generated and experimental data.
The numerical data provides an opportunity to quantify improvements and limitations of different learning approaches, 
while the experimental data allows us to demonstrate how these approaches can be applied with success to real quantum systems, even in the presence of noise and imperfections. 
For example, we show how our approach allows us to learn the Hamiltonian and Lindblad operators of the experimental device as well as the quantum efficiency of the measurement chain, while accounting for state preparation and measurement (SPAM) errors.
Our approach is therefore closer to Hamiltonian learning~\cite{Granade_robust_2012, wiebe_hamiltonian_2014, evans_scalable_2019} than quantum process tomography where a single process or a discrete collection of processes is learned~\cite{paris_quantum_2004, cramer_efficient_2010, gross_quantum_2010, torlai_neural-network_2018, torlai_quantum_2020}.

The paper is organized as follows.
After presenting the experiment, its numerical model and the generic machine-learning model 
in~\cref{sec:Experiment,sec:NumericalModeling,sec:ML}, we present two main approaches to introduce quantum features in the ML model and show their impact on numerically generated data.
Namely, we implement a loss function that encourages quantum mechanical features in a recurrent neural network in~\cref{sec:Losses} and we tailor the architecture of the ML model based on the formalism of quantum trajectories in~\cref{sec:QTMLarchitecture}.
We then apply these techniques to experimental data in~\cref{sec:ExpResults} and analyze the parameters resulting from such device characterization approaches in~\cref{sec:Params}.

\begin{figure}[t]
	\centering
    \includegraphics[width=1\columnwidth]{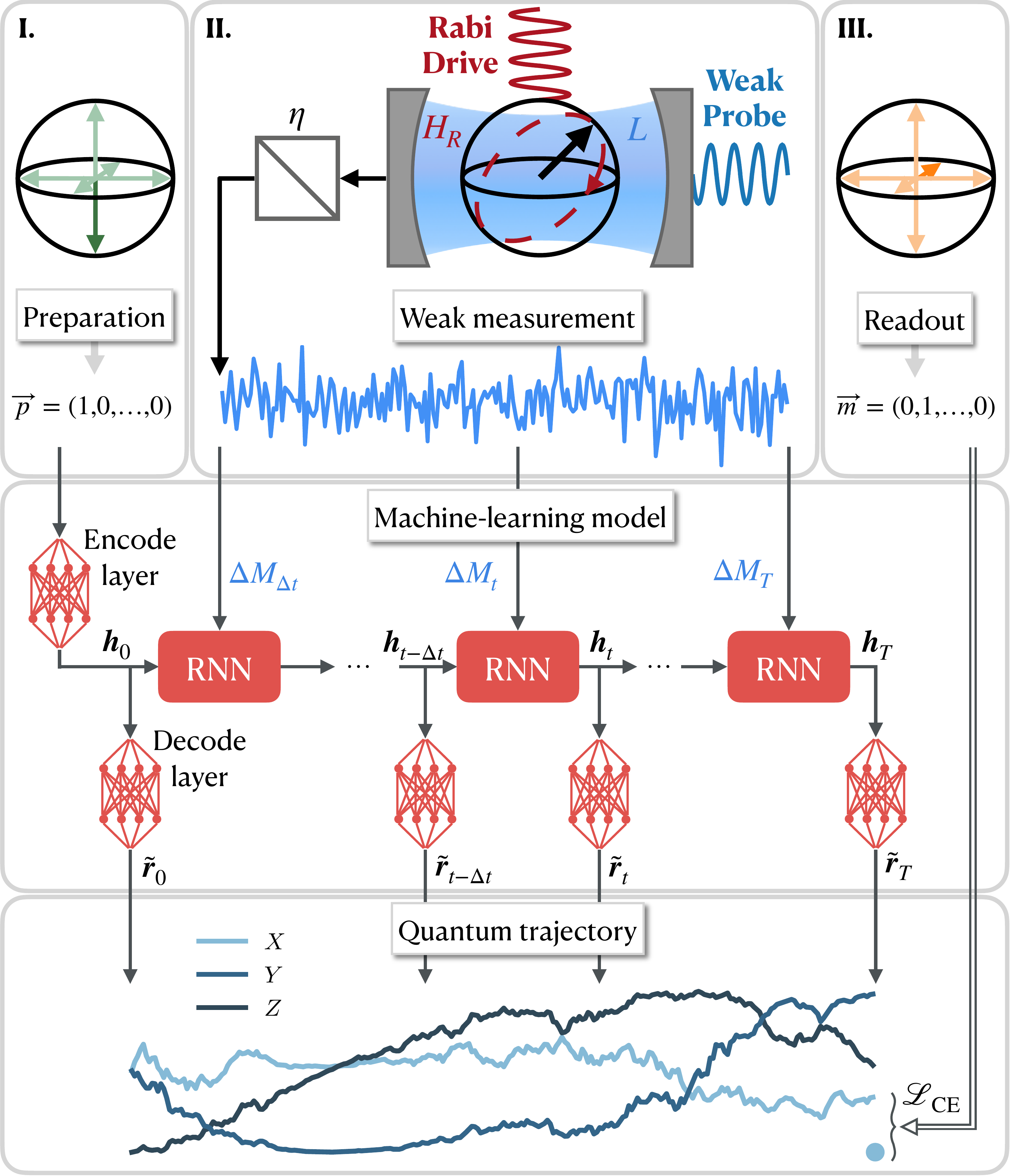}
	\caption{Schematic overview of the experiment and learning task.
	The weak measurement of a transmon qubit under the action of a resonant Rabi drive produces time-series data $\Delta M_t$.
	This information is complemented by data from state preparation and a final projective readout of the qubit, constituting the inputs to the ML model.
	The model is composed of a recurrent neural network (RNN) unit augmented by encode and decode neural layers.
	The role of these layers is to directly relate the hidden state of the network $\bm{h}_t$ with the state $\tilde{\bm{r}}_t$ of the qubit at time~$t$.
	Distinct realizations of the experiment are labeled by the outcome of the projective readout, illustrated by a light blue dot.
	The generic ML model is trained to minimize the distance between the final probability distribution associated with $\tilde{\bm{r}}_{T}$ and the actual readout outcome for an ensemble of trajectories, as quantified by the cross entropy loss $\mathcal{L}_\mathrm{CE}$.
	Relevant physical information about the qubit dynamics can be extracted from the output quantum trajectories.
	}
	\label{fig:Overview}
\end{figure}

\section{Experiment}\label{sec:Experiment}
As schematically illustrated in~\cref{fig:Overview}, the experiment consists of a superconducting transmon qubit which is continuously probed by a weak dispersive measurement via a microwave cavity.
Together with this weak measurement tone, an additional continuous tone at the qubit transition frequency produces Rabi oscillations of the qubit.
The interaction-picture Hamiltonian describing the system is $H=H_\mathrm{int}+H_R$, where~\cite{blais_RMP_2020}
\begin{align}
    H_\mathrm{int} &= \frac{\hbar \chi}{2} a^\dag a \sigma_z, \label{eq:H_int} \\
    H_R &= \frac{\hbar \Omega_R}{2} \sigma_x. \label{eq:H_R}
\end{align}
Here, $\chi$ is the dispersive qubit-cavity coupling, $\Omega_R$ the Rabi frequency, $a^\dag$ ($a$) the creation (annihilation) operator for the cavity mode, and $\sigma_{x,y,z}$ are the qubit's Pauli operators.
In our experiment, we have $\Omega_R/2\pi=0.222$~MHz, $\chi/2\pi=-0.47$~MHz, and cavity linewidth $\kappa/2\pi=1.56$~MHz. 

The experiment is realized in three essential steps.
The qubit is first 
prepared in one of the six cardinal points of the Bloch sphere $(\ket{0}, \ket{1}, \ket{\pm}, \ket{\pm i})$.
A weak measurement tone is then applied at the cavity frequency of $6.679$~GHz to continuously probe the qubit state in the $\sigma_z$ basis 
for a variable time $T$ all the while 
a microwave drive applied at the qubit frequency of $5.473$~GHz drives Rabi oscillations. 
Finally, at the end of the prescribed evolution time, a strong projective 
measurement of one of the qubit operators $\sigma_{x,y,z}$ is performed. 
After being amplified 
and digitized, the weak measurement signal takes the form of a voltage time-series $\Delta M_t$, whereas the strong readout signal is integrated and discriminated to yield a single binary output $Y\in\{+1,-1\}$.
In the machine-learning trainings, this final projective measurement is used to quantify the accuracy of the ML model in learning the quantum trajectories of the monitored qubit.

In order to build the dataset used to train our machine-learning models, we perform a total of 3.5 million shots of the three-step experiment. 
For each of the 18 possible preparation-measurement setups (6 preparations and 3 strong-measurement axes), the weak-measurement records are acquired for different total evolution time $T$ between $\SI{0}{\micro s}$ (corresponding to a projective readout immediately following the preparation) and $\SI{8}{\micro s}$ in time steps of $\Delta t=\SI{0.04}{\micro s}$. 
The resulting dataset formed by the weak measurements, together with their associated preparation and readout, is then divided into training, validation and test sets with proportions of 0.75, 0.20 and 0.05 respectively.

\section{Numerical modeling}\label{sec:NumericalModeling}
Single realizations of the experiment can be described by the formalism of quantum trajectories~\cite{wiseman_quantum_2010}.
A quantum trajectory fully captures the evolution of a qubit state $\rho$ over time.
Trajectories can be extracted from the continuous weak measurement by integrating a stochastic master equation (SME) of the form~\cite{gambetta_quantum_2008}
\begin{align}
\begin{split}
    \mathrm{d}\rho_t = & -i \left[H_R, \rho_t\right] \mathrm{d}t + \mathcal{D}[L]\rho_t \mathrm{d}t \\
    & + \sqrt{\frac{\eta}{2}}\left(\mathcal{H}[L]\rho_t \mathrm{d}W_t^\mathrm{I} +  \mathcal{H}[-iL^\dag]\rho_t \mathrm{d}W_t^\mathrm{Q}\right).
    \label{eq:SME}
\end{split}
\end{align}
In this expression, the Lindblad operator $L=\sqrt{\Gamma_d/2}\,\sigma_z$ describes the weak-measurement backaction on the qubit with $\Gamma_d$ the measurement-induced dephasing rate, which is a function of the weak measurement power, and $\eta\leq1$ is the quantum efficiency of the 
measurement chain (including signal amplification and detection)~\cite{blais_RMP_2020}.
Moreover, the independent, Gaussian-distributed Wiener increments $\mathrm{d}W_t^q$ have mean $E(\mathrm{d}W_t^q)=0$ and variance $\operatorname{Var}(\mathrm{d}W_t^q)=\mathrm{d}t$, for both quadrature $q\in\{\mathrm{I,Q}\}$ of the weak measurement signal at every time $t$.
$\mathcal{D}[L]\rho=L\rho L^\dag - \frac{1}{2}\left(L^\dag L\rho + \rho L^\dag L \right)$ is the standard dissipator 
and $\mathcal{H}[L]\rho=L\rho +\rho L^\dag - \rho \operatorname{Tr}\rho(L+L^\dag)$ is the measurement superoperator describing the backaction of the weak measurement on the quantum state~\cite{gambetta_quantum_2008}.
In~\cref{eq:SME}, we have assumed 
that both quadratures of the measurement record are monitored, something which is known as heterodyne detection~\cite{blais_RMP_2020}.
We have also assumed that the cavity-mode dynamics were much faster than that of the qubit, allowing us to adiabatically eliminate the cavity degrees of freedom from the SME~\cite{Hutchison2009}.

Experimentally, one has access to the weak-measurement record ${\Delta M^q_t}$ corresponding to a finite sampling in time of the otherwise infinitesimal signal ${\mathrm{d}M^q_t}$.
These stochastic variables are related to the Wiener increments through~\cite{gambetta_quantum_2008}
\begin{equation}
    \mathrm{d}M_t^q = \sqrt{\frac{\eta}{2}}\mathrm{Tr}\left[\rho_t(c^q + c^{q\dag}) \right] \mathrm{d}t + \mathrm{d}W_t^q,
    \label{eq:Deterministic}
\end{equation}
where $c^q=L$ ($c^q=-iL^\dag$) for the $I$ ($Q$) quadrature.
The measured signal $\Delta M_t^q$ is composed of a deterministic portion holding information about the qubit state that scales as ${\Delta t}$ and a larger stochastic contribution (noise) of order $\sqrt{\Delta t}$.

\subsection{Numerical data}\label{subsec:NumericalData}
The numerical data is generated by integrating the SME of~\cref{eq:SME} 
using the positivity-preserving scheme introduced by \textcite{rouchon_efficient_2015} and implemented in QuTiP~\cite{johansson_qutip_2013}.
To avoid most numerical integration errors from affecting the numerical datasets, we use a small integration step of $\SI{0.001}{\micro s}$ and then coarse grain the weak-measurement results to match the experimental time steps of $\Delta t=\SI{0.04}{\micro s}$. 
In this way, 1.75 million trajectories are generated and divided into training, validation and test sets with proportions of 2/3, 1/6 and 1/6 respectively.
Importantly, we generate experimentally-realistic data by using the parameters obtained from independent experimental calibration of the superconducting device. 
Below, we refer to these as the true parameters $\{\Omega_R,\Gamma_d,\eta\}$ of our physical model given by~\cref{eq:H_int,eq:H_R,eq:SME}.
Using only realistic data allows us to transparently extend the machine-learning improvements drawn from the synthetic results to actual experimental settings.

\section{Machine learning}\label{sec:ML}
Finding a physical model accurately capturing the dynamics of a quantum device and precisely calibrating all the parameters involved in the model is nontrivial and requires extensive quantum measurements and computational resources~\cite{eisert_quantum_2020}. 
However, evaluating the accuracy of a given model at predicting measurement outcomes is relatively simple.
Thus, the task of finding a good model to characterize observed quantum dynamics naturally lends itself to a machine-learning-based approach, where the accuracy of the trainable model can be optimized on a given dataset.
In this context, neural networks are particularly attractive as they can serve as universal function approximators~\cite{Goodfellow-et-al-2016}.
Specifically, recurrent neural network (RNN) architectures can preserve local time correlations~\cite{lipton_critical_2015} that are present in the weak-measurement time-ordered data.
This architecture is therefore a good candidate for the task at hand.

\subsection{Recurrent Neural Network}\label{subsec:RNN}
\Cref{fig:Overview} shows the structure of our machine-learning model.
The RNN consists of a unit cell that is repeated at every new input $\Delta M_t$ of the time-series data, producing an output $\bm{h}_t$ known as the hidden state.
This hidden state is then combined with the next time-series input $\Delta M_{t+\Delta t}$, allowing information to propagate through the sequence and weight on the outputs at future times. 
Here, we use a RNN architecture known as gated recurrent unit (GRU)~\cite{cho_learning_2014}.
Further details on the structure of our RNN implementation can be found in~\cref{appendix:NN}.
To relate the hidden state of the neural network $\bm{h}_t$ to the targeted state of the qubit $\bm{r}_t$, we complete our ML model with two feed-forward neural layers named \textit{encode} and \textit{decode} layers.
The encode layer maps a ``one-hot'' encoded preparation vector $\vec{p}$ to the initial hidden state $\bm{h}_0$ of the RNN.
The decode layer transforms a hidden state $\bm{h}_t$ to a prediction for the qubit state $\tilde{\bm{r}}_t=(\widetilde{\langle\sigma_x\rangle}_t,\widetilde{\langle\sigma_y\rangle}_t,\widetilde{\langle\sigma_z\rangle}_t)$. 

The machine-learning task then consists of inferring an accurate quantum state for the qubit at every time $t$ during the evolution, given a selected preparation $\vec{p}$ and the acquired weak-measurement time-series $\Delta M_t$.
The training label corresponds to a one-hot encoded vector $\vec{m}$ of the projective measurement outcome $Y\in\{+1,-1\}$ for the measured operator $\sigma_{x,y,z}$.
This label $\vec{m}$ represents a single bit of information about the qubit state acquired from the projective measurement at the very end of the weak measurement.
In other words, the task amounts to reconstructing the quantum dynamics of the qubit by learning its quantum trajectory $\tilde{\bm{r}}_t$, without having explicit access to the true quantum trajectory ${\bm{r}}_t$.

\subsection{Training}\label{subsec:Training}
Because in the laboratory we do not have access to $\bm{r}_t$, the task at hand is inherently difficult.
What is accessible experimentally are the coarse-grained results of the continuous weak measurement of the observable $\langle\sigma_z\rangle$, which are obscured by noise, as well as the final projective measurement outcome.
Importantly, we only train our models on this data in this work, since it represents all that is realistically available in experiments.
This is in contrast with other work where the training is based on the assumed knowledge of perfect, or slightly noisy, quantum states throughout 
the evolution~\cite{Mazza-Generators-2021, lohani_machine_2020-1, xu_neural_2018, quek_adaptive_2018}.
Although this data can in principle be acquired, it would require impractically many 
quantum process tomography experiments.
It is also worth pointing out that the learning task at hand is drastically different to other non-quantum ML approaches to solve ordinary or stochastic differential equations~\cite{chen_neural_2019,li2020scalable_PMLR}, where informationally-complete observations of the state can be made at intermediate times during the evolution. 

The training of our ML models is done by computing the gradients of a loss function $\mathcal{L}$ and updating the neural network parameters using back-propagation and gradient descent~\cite{Goodfellow-et-al-2016}.
Such a training is done using either experimental or numerically generated weak-measurement data.
In both cases, the model is incrementally updated based on its performance on the training set until the performance on the validation set starts to degrade, which indicates overfitting.
We then verify and compare the performances of the different models on the test set.

\section{Loss functions}\label{sec:Losses}

\begin{figure*}[t]
	\centering
    \includegraphics[width=2\columnwidth]{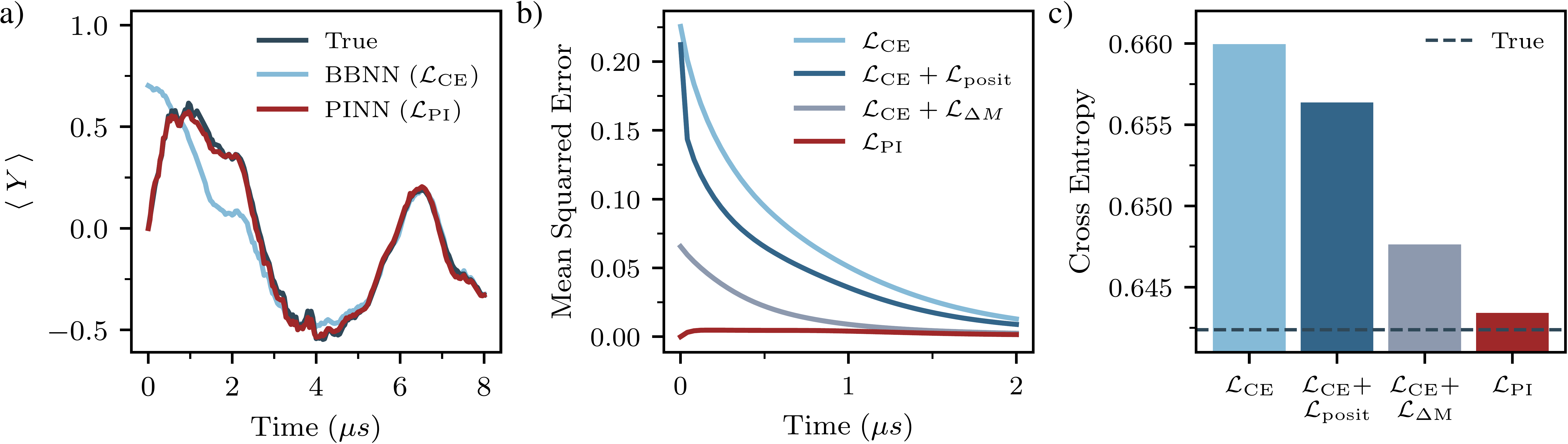}
	\caption{Including 
			quantum mechanics in the training loss function of RNN models.
	        a) Qualitative representation of the ability of a black-box neural network (BBNN) and a physics-inspired NN (PINN) to reconstruct a quantum trajectory from realistically available data.
	        The results are obtained from numerical data generated using the SME of~\cref{eq:SME} (True).
	        b) Mean squared error between the true qubit Bloch vector and the one learned by a RNN trained using different loss functions.
	        Results are shown for the first $\SI{2}{\micro s}$ of the evolution and using the entire test set trajectories.
	        Note that the models are trained on weak measurement time-series with $T=\SI{8}{\micro s}$. 
	        c) Total cross entropy for simulated projective measurements along the trajectories of the test set using a RNN and the same loss functions as panel b).
	        }
	\label{fig:2}
\end{figure*}

\subsection{Cross Entropy}
In supervised learning, a black-box approach might consists in training a generic neural network using a loss function that is agnostic to any features or physics associated with the learning task.
The loss function simply measures the distance between the true observed outputs (of the labelled dataset) and the predicted outputs of the model. 
An example is the binary cross entropy, also known as negative log-likelihood, which in our case quantifies the separation between the qubit-state measurement probabilities predicted by the ML model and the observed projective-measurement outcomes~\cite{flurin_using_2020}.
The cross-entropy loss takes the form
\begin{equation}\label{eq:LossFunctionCEL}
	\begin{split}
		&\mathcal{L}_\mathrm{CE} = \\
		& -\frac{1}{N}\sum_{n=1}^N 
	\left[\frac{(1+Y_n)}{2} \log(\Pi_n^\alpha) + \frac{(1-Y_n)}{2} \log(1-\Pi_n^\alpha)\right],
	\end{split}
\end{equation}
for the pairs $(Y_n, \Pi_n^\alpha)$ of projective measurement outcome $Y_n \in \{+1,-1\}$
and the probability $\Pi_n^\alpha \in [0,1]$ predicted by the ML model of measuring the outcome $+1$ for the operator $\sigma_\alpha$.

In addition to using $\mathcal{L}_\mathrm{CE}$ 
as a training loss, we use the cross entropy to compare 
the performance of different models at predicting the experimentally observed (or simulated) measurement outcomes.
Using the cross entropy as a performance metric is motivated by the fact that, in the present case where the measurement outcomes are fixed, it is equivalent up to an additive constant to the Kullback-Leibler divergence~\cite{Goodfellow-et-al-2016}.
Moreover, using the cross entropy allows us to quantify the accuracy of the trained ML model at the level of its individual output trajectories, without having to introduce statistical errors by binning or averaging the trajectories to yield the desired metric.

\Cref{fig:2} presents the performance of RNNs trained on numerically generated data -- we come back to experimental data in \cref{sec:ExpResults}.
Panel~a) illustrates the deviations of the learned quantum trajectory (light blue line) from the true trajectory (dark blue line) for a single representative component of $\bm{r}_t$ and $\tilde{\bm{r}}_t$ sampled from the test set.
As is made clear from this example, because the model is only trained to be accurate at the very end of the quantum evolution where the projective measurement is performed, there are little variations between true and learned trajectories with the generic $\mathcal{L}_\mathrm{CE}$ black-box neural-network (BBNN) approach at late times.
However and for the same reason, the learned quantum trajectories deviate from the true trajectories at early times.
This observation is made more quantitative in~\cref{fig:2}b), which shows the mean squared error as a function of time between the true trajectories and those learned using the cross-entropy loss (light blue line).
The corresponding cross entropy between the generated trajectories of the entire test set and the RNN-learned ones is, moreover, shown in~\cref{fig:2}c).

In short, a RNN trained with the loss function $\mathcal{L}_\mathrm{CE}$ is relatively accurate at predicting the measurement outcomes of unseen data, 
but fails to precisely capture quantum dynamics during the entire evolution.
Extracting information about the device parameters with this approach is therefore expected to lead to flawed results.

\subsection{Physics-Inspired Loss Functions}

\begin{figure}[t]
	\centering
    \includegraphics[width=\columnwidth]{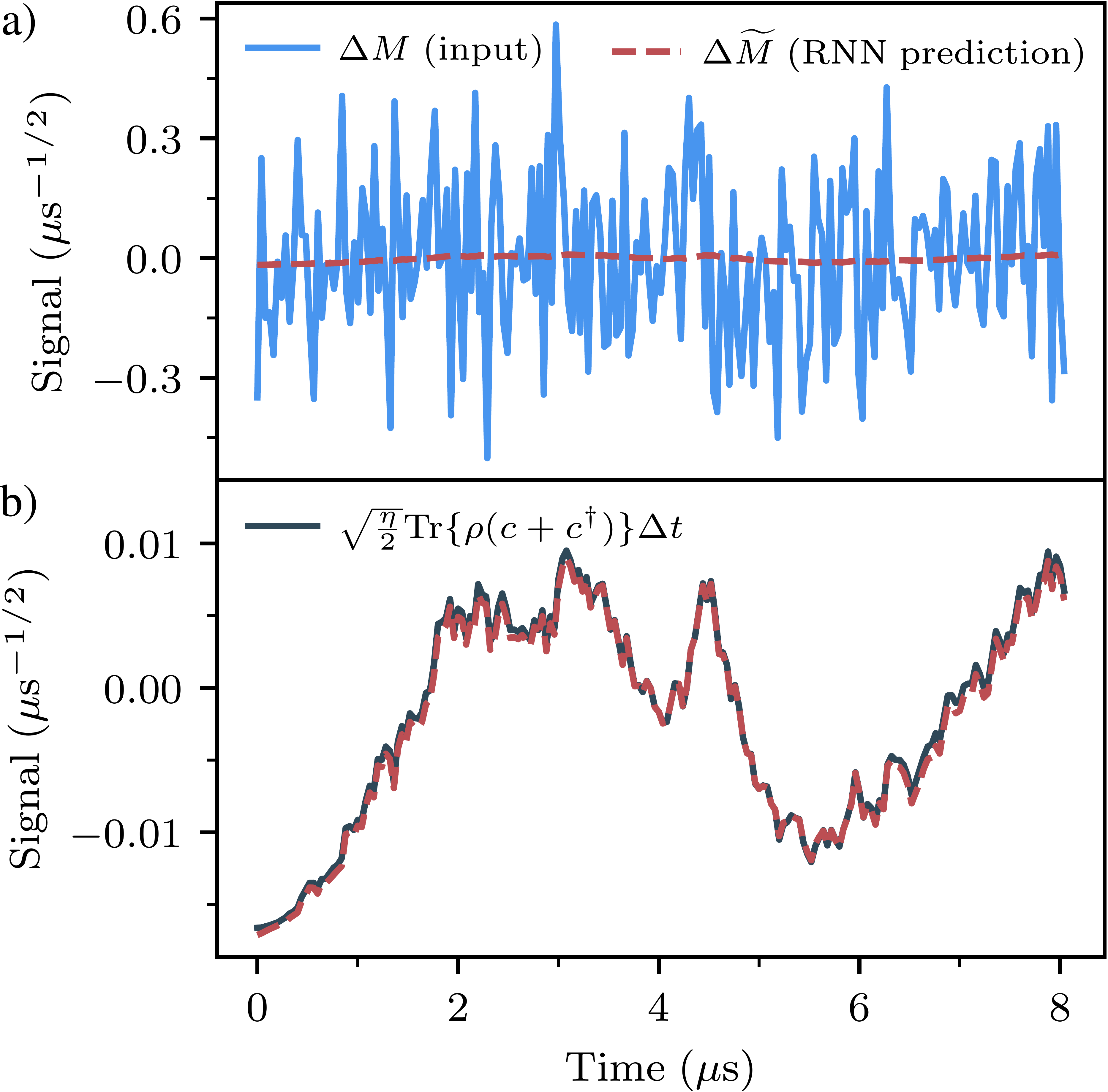}
	\caption{Finding the deterministic information of a noisy signal using the physics-inspired loss of~\cref{eq:LossFunctionPredict}.
	a)~Weak-measurement input signal (blue) and recurrent neural-network prediction of the same signal (dashed red) for a given quantum trajectory of the test set. 
	The input signal is composed of information about the qubit state, through the expectation value of the $(c+c^\dag)$ operator, but it is dominated by quantum noise, see~\cref{eq:Deterministic}.
	b) Zoom-in of the RNN prediction and comparison to the true value of the deterministic information about the qubit state (dark blue).
	Results are obtained from numerically generated data.
	}
	\label{fig:PredLoss}
\end{figure}

To improve the situation, we now expend the loss function with terms penalizing learned trajectories which do not respect specific quantum mechanics features.
Such regularization of the training allows the RNN to explore the optimization space more efficiently~\cite{raissi_physics-informed_2019}. 

Our first physics-inspired loss is the \textit{positivity} loss, which is aimed at penalizing ML outputs corresponding to unphysical qubit states living outside of the Bloch sphere.
It takes the form
\begin{equation}
    \mathcal{L}_\mathrm{posit} = \frac{1}{N(N_t+1)}\sum_{t=0}^{N_t} \sum_{n=1}^N \mathrm{ReLU}\left(\left|\tilde{\bm{r}}_{t,n}\right|^2 - 1 \right),
    \label{eq:LossFunctionPosit}
\end{equation}
where $\mathrm{ReLU}(x)=\operatorname{max}(0,x)$ is the rectified linear unit and, as above, $\tilde{\bm{r}}_t=(\widetilde{\langle\sigma_x\rangle}_t,\widetilde{\langle\sigma_y\rangle}_t,\widetilde{\langle\sigma_z\rangle}_t)$ is the RNN prediction of the qubit state.
We note that a BBNN tends to learn states that violate this positivity constraint at early times.

The second loss term proposed is the \textit{preparation} loss which is intended to make the RNN predictions accurate at the beginning of the quantum trajectory by using our knowledge of the qubit's initial state. 
This loss is expressed as
\begin{equation}
    \mathcal{L}_\mathrm{prep} = \frac{1}{N}\sum_{n=1}^N \left|\tilde{\bm{r}}_{0,n} - \bm{r}_{0,n}\right|^2,
    \label{eq:LossFunctionPrep}
\end{equation}
where, when dealing with experimental data, we can set the targeted preparation states $\bm{r}_{0}$ to be consistent with state preparation and measurement (SPAM) errors.
To do so, we consider the subset of data where the projective readout immediately follows the preparation (i.e. the data with a weak-measurement time $T=\SI{0}{\micro s}$).
The states $\bm{r}_{0}$ are then inferred by extracting the expectation value of the $\sigma_{x,y,z}$ operators on the prepared cardinal states of the Bloch sphere, thus effectively carrying out quantum state tomography of these initial states~\cite{paris_quantum_2004}.
This simple pre-processing of the dataset based on our understanding of the dynamics then allows us to train the ML model to output the best initial states given the same measurement data.

The third physics-inspired loss that we introduce is the \textit{prediction} loss.
The objective is to force the model to use directly the deterministic information about the qubit state that is present in the time-series data.
To do so, we add an output $\Delta\widetilde{M}_t^q$ to the RNN model that serves as a prediction for the next input of the time-series weak-measurement data $\Delta M_{t+\Delta t}^q$. 
The prediction loss is implemented as a 
mean squared error
\begin{equation}
    \mathcal{L}_{\Delta M} = \frac{1}{N_t N}\sum_{t=0}^{N_t-1}  \sum_{n=1}^N \left(\Delta\widetilde{M}_{t,n}^q - \Delta M_{t+\Delta t,n}^q\right)^2,
    \label{eq:LossFunctionPredict}
\end{equation}
that pushes the model to make predictions about the next weak measurements that are as accurate as possible.
Note that the mean squared error is equivalent to a cross-entropy loss for a Gaussian random variable like $\Delta M_{t}^q$.

Adding these different contributions, our physics-inspired loss function takes the form
\begin{equation}
    \mathcal{L}_\mathrm{PI} = \mathcal{L}_\mathrm{CE} + w_\mathrm{posit}\mathcal{L}_\mathrm{posit} + w_\mathrm{prep}\mathcal{L}_\mathrm{prep} + w_\mathrm{\Delta M}\mathcal{L}_\mathrm{\Delta M}, \label{eq:LossFunction4L}
\end{equation}
where $w_j$ are relative weights that can be optimized similarly to other hyperparameters of typical NN trainings.
As illustrated in \cref{fig:2}, the trajectories resulting from training the RNN with the physics-inspired loss function $\mathcal{L}_\mathrm{PI}$ are significantly more accurate than those obtained using only a physics-agnostic $\mathcal{L}_\mathrm{CE}$.
This is illustrated qualitatively in \cref{fig:2}a) where the red line correspond to a representative result obtained using \cref{eq:LossFunction4L}.
In contrast to the results obtained with $\mathcal{L}_\mathrm{CE}$ (light blue line), the predicted trajectory now matches the true trajectory (dark blue line) significantly better at early times.
Panels b) and c) show the importance of the different physics-inspired loss terms by presenting the mean squared error and the cross entropy, respectively, for RNNs trained on different subsets of $\mathcal{L}_\mathrm{PI}$.

\Cref{fig:PredLoss} illustrates how $\mathcal{L}_\mathrm{PI}$, and specifically the prediction loss term $\mathcal{L}_{\Delta M}$, helps the RNN learn significantly better quantum trajectories.
Panel a) shows a representative example of an input weak-measurement signal $\Delta M$ that was numerically generated (light blue) and the prediction of the same signal $\Delta\widetilde{M}$ (dashed red line) from a RNN trained using \cref{eq:LossFunction4L}. 
We see that these two curves do not directly correspond, which is expected from the fact that $\Delta M$ is mostly composed of random noise (proportional to $\sqrt{\Delta t}$) coming from the Wiener increments in~\cref{eq:Deterministic}.
\Cref{fig:PredLoss}b) shows the agreement between the deterministic portion of this input signal (dark blue) and the same RNN prediction --- notice the change of scale.
This result shows that our physics-inspired RNN model is able to infer the deterministic information (proportional to ${\Delta t}$) hidden in the noisy weak-measurement signal, even if this relevant information is at least an order of magnitude smaller than the noise for our experimentally-realistic parameters.
Interestingly, the ML model is then able to use the information associated with this weak-measurement prediction in order to output quantum trajectories that are significantly more accurate throughout the entire quantum evolution.

\section{Quantum-tailored ML architecture}\label{sec:QTMLarchitecture}
Thus far, we have used a generic RNN and shown how leveraging prior knowledge about the structure of a quantum problem allowed us to learn more accurate dynamics while using the same input data.
In this section, we follow this intuition further by incorporating features of quantum mechanics in the machine-learning model architecture itself.
To do so, we use a physically interpretable model to learn a useful characterization of the device from the weak-measurement data.

The idea consists of directly using our physical model of the quantum dynamics, given by~\cref{eq:SME}, as the trainable model.
This approach makes use of the direct relation between the hidden state of the RNN and the quantum state of the device.
Since all measurement predictions are derived from the hidden state of the RNN, the vector corresponding to the hidden state holds an overcomplete representation of the quantum state, and the layers that translate it into Pauli-measurement probabilities can be used to map this representation into the usual density-matrix representation: $\bm{h}_t\to\rho_t$.
As such, the function $\mathcal{F}$ that the RNN is set out to learn is the one updating the quantum state given weak measurement data
\begin{align}
    \bm{h}_{t+\Delta t} &= \mathcal{F}_\mathrm{RNN}(\bm{h}_t, \Delta M_t^q),\\
    \begin{split}
        \rho_{t+\Delta t} &= \mathcal{F}_\mathrm{SDE}(\rho_t, \Delta M_t^q) \\
        &= \rho_t + \Delta \rho(\rho_t, \Delta M_t^q),
        \label{eq:SDE-SME}
    \end{split}
\end{align}
where the last equation is a discretized integration of~\cref{eq:SME}.
Thus, we can use a stochastic differential equation (SDE) integrator, with a 
set of free parameters, as our learnable model implementing $\mathcal{F}_\mathrm{SDE}$.
We set those free parameters to be the frequencies, rates and operators that describe the experimental setup, such as $\{H_R,L,\eta\}$ that appear in~\cref{eq:SME}.
With this choice, learning accurate quantum trajectories from weak-measurement data becomes a perfectly interpretable device characterization.

We implement the SDE integrator based on the Milstein method~\cite{kloeden_numerical_1994} 
using an auto-differentiable library. 
This implementation allows the model to operate on a GPU and to be trained efficiently using back-propagation in the same way as generic neural networks~\cite{Goodfellow-et-al-2016}.
In the following, we will refer to this machine-learning approach as SDE learning.
We note that the use of a differential equation integrator as a trainable model can be straightforwardly extended to other parameter-estimation problems described by a differential equation.
As such, the authors of Ref.~\cite{krastanov_unboxing_2020} independently developed a similar learning approach for control problems, although they only analyze the performance on numerical data with high temporal resolution.
We emphasize that realistic experimental restrictions, such as the size of the time increment $\Delta t$,
have important consequences on the performance of these learning schemes, which we address in \cref{appendix:CoarseGraining}.

Using SDE learning on numerically generated data, we have found that this approach is able to reconstruct the true quantum trajectories up to numerical-integration errors (not shown).
Here, errors are due to the finite and realistic time step of the weak-measurement data, which is set to $\Delta t=\SI{0.04}{\micro s}$. 
The SDE model based on~\cref{eq:SME} outperforms the RNN models at the task of learning the correct system dynamics, achieving a total mean squared error of $6.3\times 10^{-5}$ on the trajectories of the test set.
In comparison, the physics-inspired RNN gives an error of $5.1\times 10^{-3}$ when trained on the same data. 
Importantly, the SDE-learning approach also directly outputs the device parameters and the generators of the quantum dynamics.
The characterization of the device resulting from this approach is further discussed in~\cref{sec:Params}.
We note that we can also account for SPAM errors in the SDE-learning approach by learning the preparation and projective measurement maps from the initial state tomography subset data, similarly to what can be done with the preparation loss. 

While the excellent performance of the SDE learning is to be expected for numerically generated data, the next section explores the application of this approach to experimental data where the model of~\cref{eq:SME} does not capture all of the physical dynamics present in the weak measurements. 

\begin{figure*}[t]
	\centering
    \includegraphics[width=2\columnwidth]{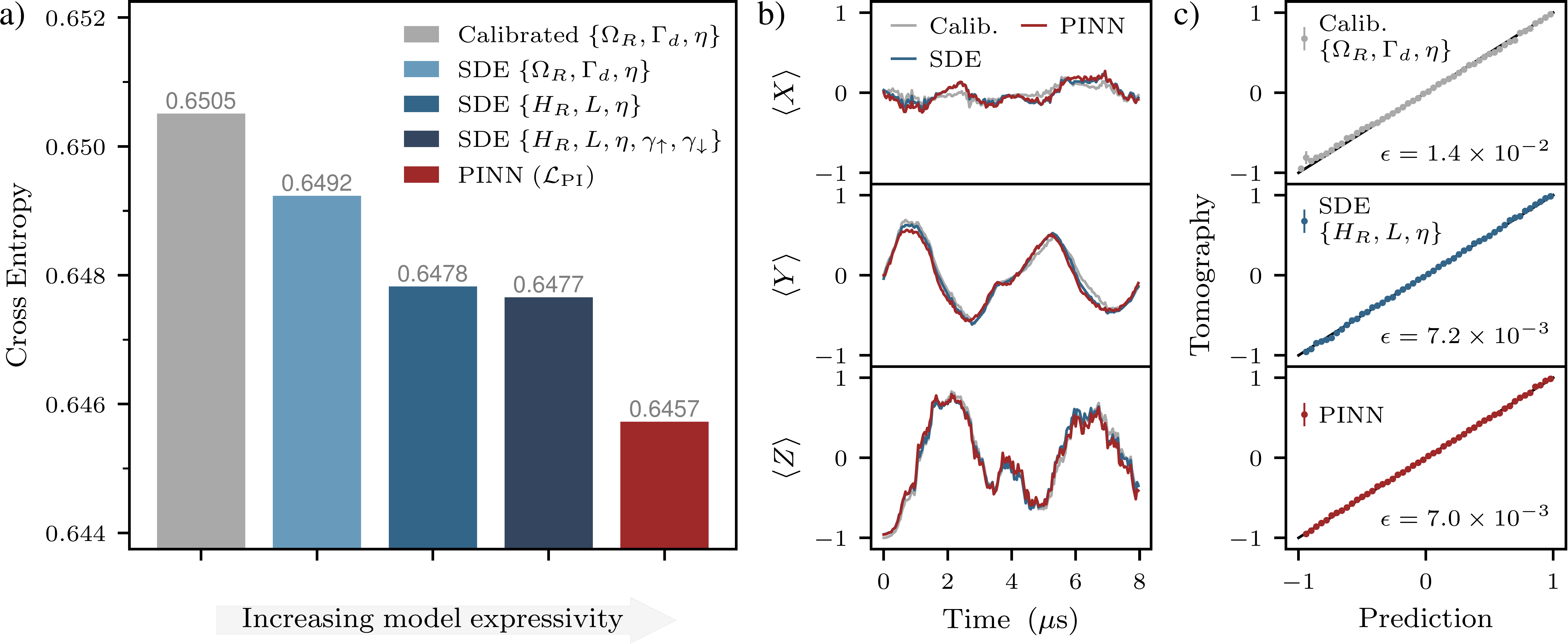}
	\caption{Performance of the 
	ML models to predict experimental outcomes. 
	a) Cross entropy measure on the test dataset for the physical model of~\cref{eq:SME} using independently calibrated parameters (gray) or learning the parameters from the weak measurement data in a SDE model (light blue), additionally learning the Hamiltonian and Lindblad operators (blue) and qubit relaxation rates (dark blue).
	The performance of a physics-inspired recurrent neural network (PINN) trained with the loss function of~\cref{eq:LossFunction4L} is also shown (red).
	b) A quantum trajectory reconstructed from an input weak measurement time-series sampled from the test dataset for three distinct models of panel a).
	c) More visual comparison of the model predictions with the observed tomography measurements for the same three models.
	The three tomography axes are used and averaged over.
	Also presented is the root mean squared error $\epsilon$ of the model predictions for all points extracted from the validation dataset (total of $0.68$~M trajectories).
	The error bars are given by the standard deviation of the binomial distribution of model predictions for each point.
	}
	\label{fig:ExperimentalRes}
\end{figure*}

\section{Experimental Results}\label{sec:ExpResults}
We now move from numerically generated data to experimental data obtained from measurements on a superconducting qubit. 
Because we have already limited the training to experimentally accessible quantities, there are no differences in the training process. 
However, since the true quantum trajectories are now unknown and the experimental data is affected by nonidealities and noise beyond what can be captured with the physical model of~\cref{eq:SME}, quantifying the degree of success of the training is more difficult in this case. 
For this reason, we use the cross entropy to quantify the ability of the trained models to capture the qubit dynamics based on their predictions for the projective measurement acquired at the end of every experiment. 

\Cref{fig:ExperimentalRes}a) shows the cross entropy obtained using three different approaches: the physical model of \cref{eq:SME} with independently calibrated parameters (gray), SDE learning (blue), and physics-inspired RNN trained using $\mathcal{L}_\mathrm{PI}$ (red).
As an illustration of the outcome of training our ML models on experimental data, \cref{fig:ExperimentalRes}b) shows a representative quantum trajectory reconstructed with these three approaches while using the same input weak-measurement data.
Comparing the gray and left-most 
blue bars in \cref{fig:ExperimentalRes}a), we see that the SDE-learning approach allows us to learn physical parameters that describe the experimental outcomes significantly better than inferring these same parameters, $\{\Omega_R,\Gamma_d,\eta\}$ appearing in~\cref{eq:SME}, using traditional calibration experiments~\cite{murch_observing_2013}.
Part of this discrepancy between the calibrated model and the learned SDE model can be attributed to calibration errors and experimental drifts which are likely to occur during the experiment.
The better performance of the SDE learning demonstrates the ability of this approach to learn an accurate physical model of the quantum device.

\Cref{fig:ExperimentalRes}a) also compares the performance of three SDE models (blue bars) that have an increasing number of degrees of freedom from left to right.
The left-most blue bar is obtained by learning the parameters $\{\Omega_R,\Gamma_d\}$ of the Hamiltonian $H_R=\Omega_R/2 \sigma_x$ and Lindblad operator $L=\sqrt{\Gamma_d/2} \sigma_z$, together with the quantum efficiency $\eta$.
A significant gain in accuracy can be obtained by leaving unconstrained the form of the Hamiltonian and Lindblad operators and learning these full operators $\{H_R,L\}$ (middle blue bar) directly from the data.
Finally, taking advantage of the flexibility of the SDE-learning approach, we can further enrich the model that is learned, for example by adding to \cref{eq:SME} qubit excitation and relaxation described by the additional dissipators $\gamma_{\uparrow}\mathcal{D}[\sigma_{+}]\rho_t\mathrm{d}t + \gamma_{\downarrow}\mathcal{D}[\sigma_{-}]\rho_t\mathrm{d}t$.
As can be seen from the right-most blue bar, the gain in predictive power is, however, marginal reflecting the fact that this particular addition is not capturing a meaningful contribution to the dynamics of our physical system.
As presented with these three models, the ability to directly learn SDEs makes it straightforward to define nested physical models with growing parameter sets and use model selection to determine the best physical description that balances between predictive power and economy in parameters~\cite{Wilks-likelihoodratio-1938}.

The performances of the SDE models are further compared in~\cref{fig:ExperimentalRes}a) with a physics-inspired RNN trained on the same weak-measurement data (red bar).
The RNN has a greater expressivity and is able to achieve a lower cross entropy since it is free from the constraints of the SDE-learning model. 
This result shows that some qubit dynamics are captured by the RNN but not by the presented SDE models. 
An advantage of SDE learning remains, however, in that it is a physically interpretable approach which accuracy can be improved by adding dynamics to the physical model describing the device.
For example, it would be possible to include the readout cavity mode in the SDE model in order to characterize the non-Markovian qubit dynamics that are likely to be present in the experiment~\cite{gambetta_quantum_2008}.

\begin{figure}[t]
	\centering
    \includegraphics[width=1\columnwidth]{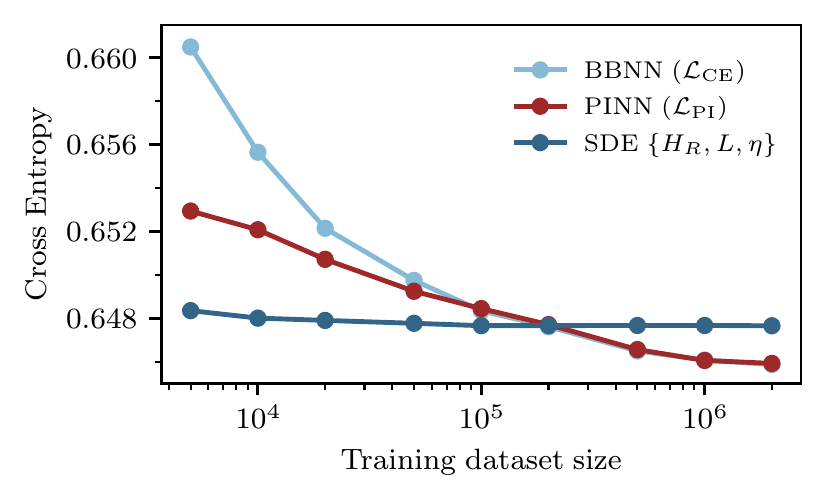}
	\caption{Model performance as measured by the cross entropy on the test set as a function of the number of experimental weak-measurement training samples.
	The same RNN model is trained on the generic loss function~\cref{eq:LossFunctionCEL} in light blue and on the physics-inspired loss function \cref{eq:LossFunction4L} in red.
	The SDE-learning model in blue has the structure of~\cref{eq:SME} with the free parameters identified in the braces.
	}
	\label{fig:Scaling}
\end{figure}

Another advantage of SDE learning is that it requires significantly less experimental training data.
This is illustrated in \cref{fig:Scaling} which shows the cross entropy as a function of the size of the training dataset.
We see that SDE learning (blue line) requires more than a order of magnitude fewer training samples than RNNs (light blue and red lines) to reach the same prediction accuracy.
Of course, as already pointed out, the finite expressive power of the SDE model limits its maximal accuracy, which reaches a plateau while the RNN performance's continue to improve with millions of training samples.
Additionally, \cref{fig:Scaling} shows that the physics-inspired loss $\mathcal{L}_\mathrm{PI}$ allows the RNN to learn more accurate quantum trajectories with fewer data than using the physics-agnostic loss~$\mathcal{L}_\mathrm{CE}$. 
Given enough training data, the black-box RNN is however able to infer the known quantum features imposed to the physics-inspired model in order to perform equally well on the cross entropy, albeit this BBNN outputs non-positive quantum trajectories.

To illustrate the prediction accuracy of the SDE learning and physics-inspired RNN in a less rigorous but more intuitive way than the cross entropy, \cref{fig:ExperimentalRes}c) presents the measure of self consistency used in Refs.~\cite{ficheux_dynamics_2018, flurin_using_2020}. 
This measure consists in first binning the learned trajectories predicting the same probability $\Pi_n^\alpha$ for the final measurement outcome within a small $\delta=0.04$.
We then compare this approximate prediction to the average final projective measurement outcome (i.e.~the tomography result) for the same set of trajectories. 
Both the SDE learning and RNN models are in excellent correspondence and follow the expected unit slope.
The overall agreement between the model predictions and the tomographic measurements is quantified by the root-mean-squared error $\epsilon$ weighted by the number of trajectories in each bin.
All the same, this measure is not as sensitive as the cross entropy which compares every single learned quantum trajectory to its associated projective measurement.

\section{Analysis of device parameters}\label{sec:Params}
We now study the device 
parameters learned from 
our two machine-learning approaches, namely the SDE-learning and the physics-inspired RNN.
The SDE model can be trained directly on the weak-measurement data to estimate its free and interpretable parameters.
On the other hand, the representation learned by the RNN is not as straightforward to interpret.
Here, we propose an efficient way to extract the device parameters using the SDE-learning approach on the RNN-learned quantum trajectories.

\begin{figure}[t]
	\centering
    \includegraphics[width=1\columnwidth]{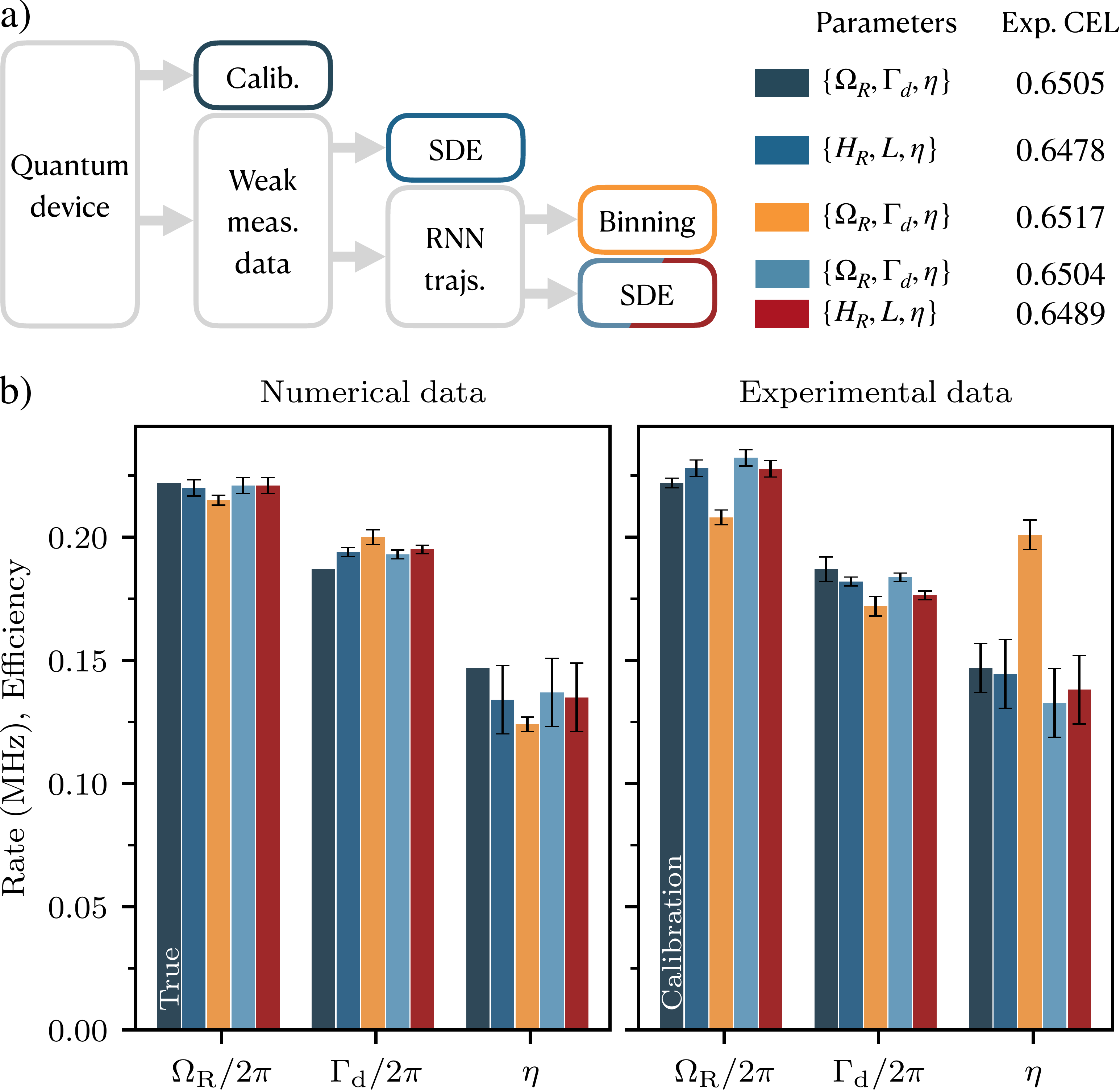}
	\caption{Characterizing quantum dynamics from weak-measurement data.
	a) Schematic of the different approaches to infer qubit parameters.
	On the rigth hand side, the cross entropy loss (CEL) of the different models on the experimental test set are presented.
	A lower CEL implies a better prediction of the actual measurement outcomes.
	The free parameters of the models are shown in braces.
	For the last three approaches, a RNN is first trained on the weak-measurement data using the physics-inspired loss function~\cref{eq:LossFunction4L}, achieving a CEL of 0.6457, and the parameters are extracted from the RNN output trajectories.
	b) Device parameters extracted from the different learning approaches on numerically generated data (left panel) and on experimental transmon data (right panel).
	Note that we consider the parameters $\Omega_R/2$~($\sqrt{\Gamma_d/2}$) to be the $\sigma_x$~($\sigma_z$) component of $H_R$~($L$).
	The error bars for the SDE approaches are extracted by comparing performance on numerical data, see~\cref{appendix:CoarseGraining} for more details.
	The error bars of the calibration and binning approaches are extracted directly from curve fits.
	}
	\label{fig:Params}
\end{figure}

An overview of the different parameter inference approaches considered in this work is presented in~\cref{fig:Params}a), along with the parameters that are learned and the cross-entropy loss achieved on experimental data.
\Cref{fig:Params}b) shows the parameters $\{\Omega_R,\Gamma_d,\eta\}$ obtained from these approaches on numerical data (left panel) and experimental data (right panel).
By comparing the darkest blue bar (true values used to generate the numerical data) to the second blue bar (SDE learning) in the left panel, we see 
that the SDE learning is able to infer the device parameters with an accuracy limited only by numerical errors introduced in the SDE integration.
As presented in~\cref{appendix:CoarseGraining}, this error is proportional to the time step $\Delta t$ which we take to be the same as for the experimental data, namely $\Delta t=\SI{0.04}{\micro s}$.
The impact of coarse-graining the weak-measurement data is 
analyzed in~\cref{appendix:CoarseGraining}.
\Cref{fig:Operators} further confirms the excellent performance of SDE learning on numerical data by showing the full Hamiltonian (top row) and Lindblad operators (bottom row) that are learned.
These Hinton diagrams directly show the learned $2\times2$ matrices.
Comparing the first two columns of this figure, we confirm that the characterization obtained on numerical data matches very well with the true values.

\begin{figure}[t]
	\centering
    \includegraphics[width=1\columnwidth]{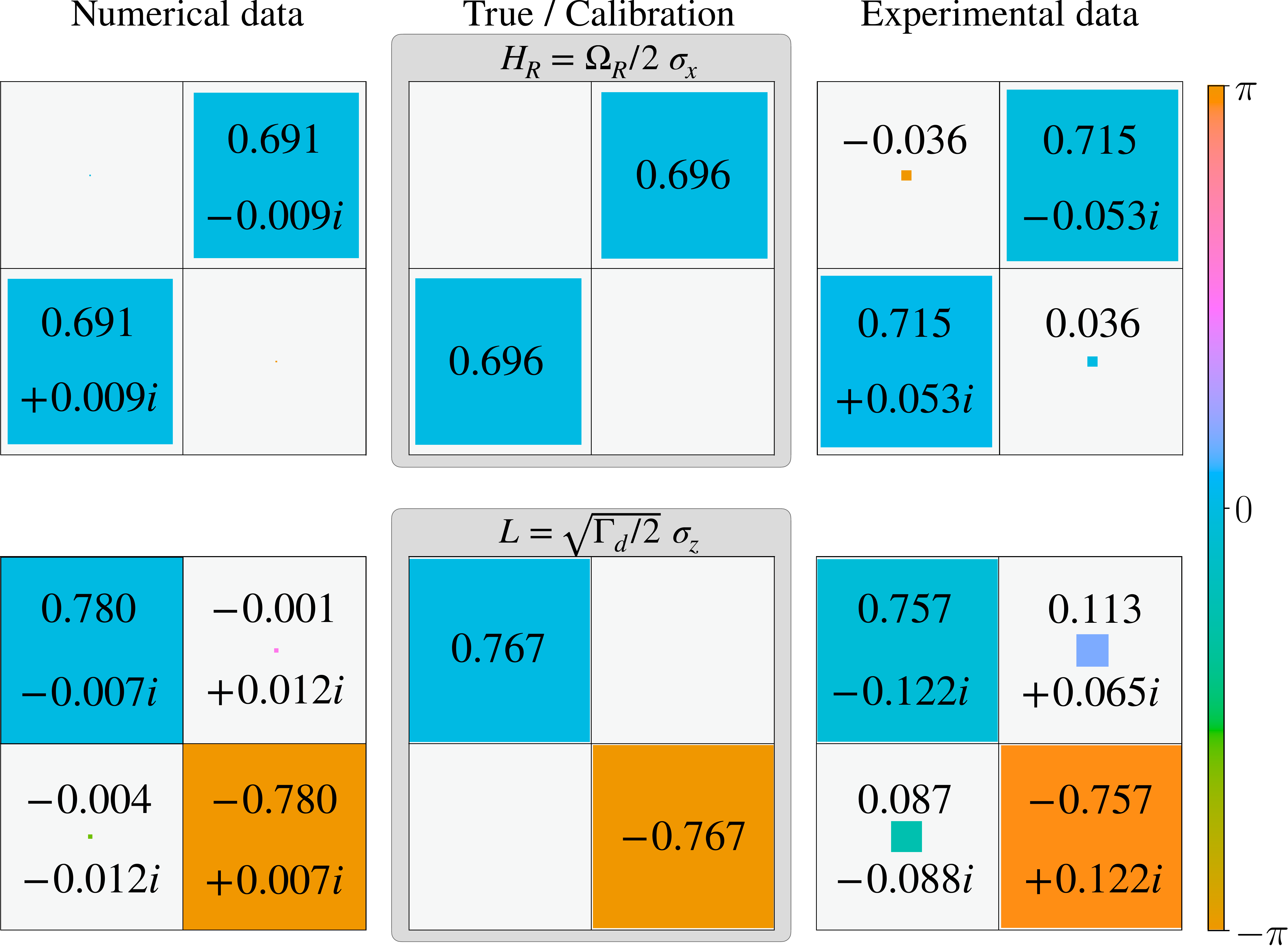}
	\caption{
	Hinton diagrams of the Hamiltonian $H_R$ (top row) and Lindblad $L$ operators (bottom row) learned by the SDE model with free parameters $\{H_R,L,\eta\}$ trained directly on the weak measurements data.
	These $2\times2$ matrix operators are presented for the training on numerical data (left column) and on experimental data (right column) in comparison to the true and calibrated operators chosen to be the same (center column). 
	The size of the colored squares is proportional to the absolute value of the complex number entry and the color is associated with its phase. 
	For clarity, values are rounded to the third decimal and null values are not shown. 
	}
	\label{fig:Operators}
\end{figure}

On the other hand, the parameters obtained from SDE learning on experimental data show larger deviation from the values obtained from independent calibration of the device, see the right panel of \cref{fig:Params}b).
The same is true for the full Hamiltonian and Lindblad operators shown in \cref{fig:Operators}; compare the last two columns.
The deviations of the inferred parameters from their ideal values are indicative of the need for a more sophisticated model~\cite{koolstra_2021}.
Further analyzing these discrepancies between the calibrated and learned models could lead to a better understanding of the qubit dynamics.
This interpretability of the SDE-learning approach constitutes one of its important advantage as opposed to black-box learning approaches.

Motivated by interpreting the model achieving the lowest cross entropy on the experimental data, we now analyze how we can physically interpret a RNN trained on the weak measurements.
The approach to do so used in Refs.~\cite{flurin_using_2020, atalaya_correlators_2019}
consists of two steps.
First, the output trajectories are binned within some small expectation value intervals $\delta$ at every time point.
Second, the mean and variance of these distributions are fitted to a SME model using least squares fits, see Refs~\cite{flurin_using_2020, atalaya_correlators_2019} for more details.
Device parameters resulting from this approach are shown in \cref{fig:Params}b) with the yellow bars.
In the limit of vanishing bin size and many trajectories in each bin this approach will recover the true device parameters.
Unfortunately, in the finite-data regime the fits are sensitive to the particular choice of bin size, with larger bin sizes discarding information due to coarse graining and smaller bin sizes suffering from reduced sampling statistics.

To improve upon this method, we employ the SDE-learning approach to interpret the learned representation of the RNN.
This is achieved by training the SDE model to output quantum trajectories that maximize the likelihood of matching the learned RNN trajectories.
This training task is much simpler than training on the projective measurement outcomes, since we now have direct access to the target quantum trajectories outputted by the RNN.
To minimize the distance between the SDE and the RNN trajectories, we use a mean squared error loss which provides good numerical convergence and is expressed as
\begin{equation}
    \mathcal{L}_\mathrm{MSE} = \frac{1}{N(N_t+1)}\sum_{t=0}^{N_t}\sum_{n=1}^{N} \left(\tilde{\bm{r}}_{t,n}^\mathrm{SDE}-{\bm{r}}_{t,n}^\mathrm{RNN}\right)^2,
\end{equation}
for $N$ trajectories with $N_t$ time steps.

As shown in~\cref{fig:Params}a) by comparing the cross entropy of the lightest blue and yellow colors, using the SDE learning allows to recover more accurate device parameters from the trained RNN than using the existing binning approach.
As opposed to the binning approach, we show in red how the SDE model can easily be extended to include more free parameters, in this case the full quantum operators $\{H_R,L\}$, in order to achieve a better device characterization. 
Moreover, the SDE-learning performs a maximum likelihood estimation of the physical parameters~\cite{krastanov_stochastic_2019}, thus making our method ideally suited to the task of interpreting a trained RNN and preferable to a binning approach.
It is also worth mentioning that the problem of characterizing the quantum dynamics can now be broken down into two independent task: while a single RNN with high expressivity can be used to learn an accurate yet opaque classical representation of the quantum dynamics, a set of SDE models can then be used to associate a physical meaning to the result and extract relevant device parameters.

\section{Conclusion}\label{sec:Conclu}
We have demonstrated that leveraging quantum mechanics in the design of a machine-learning approach makes the characterization of quantum dynamics more efficient and accurate when using limited experimentally available data.
We have analysed both numerically generated and experimental data obtained with a superconducting qubit, showing the tradeoffs between the accuracy of the learned description, its efficiency in the number of training samples required and its interpretability that allows such a description to be useful for the characterization of the quantum device. 
We have presented two main approaches to acquire an accurate heuristic of the quantum dynamics from weak-measurement data, namely i) by designing the loss function of a recurrent neural network to make use of our understanding of the quantum formalism and ii) by tailoring the structure of the machine learning model to respect the structure of the stochastic differential equation describing the device dynamics.

Overall, our results demonstrate that useful insights about the physics of quantum systems can be gained by interpreting machine-learning models trained on experimental data, thus going further than using black-box learning approaches.
In particular, the SDE-learning constitutes a promising avenue to perform realistic qubit modeling and characterization via a single weak-measurement experiment.
Notably, this approach allows us to characterize the quantum efficiency of the measurement chain $\eta$, which usually requires involved characterization~\cite{bultink_general_2018}.
The ideas developed in this work 
can be naturally extended to other quantum tasks involving, for example, noise characterization and optimal quantum control. 
Our characterization method is also advantageous for continuous quantum feedback control, where controls are applied based on continuous monitoring of the system~\cite{zhang_quantum_2017}.

An objective of the machine-learning approaches introduced in this work is to acquire an efficient representation of complex quantum devices to improve their characterization, calibration and control.
Towards this end, it will be interesting to extend the learning to more qubits and to explore the characterization of additional quantum phenomena occurring in these more complex devices, such as non-Markovian dynamics and crosstalk.
The higher training efficiency of our quantum-tailored machine-learning approaches might become especially relevant when scaling up these quantum tasks, where the limitations of available experimental data are increasingly stringent.

\begin{acknowledgments}
    We thank Simon Verret for insightful discussions.
	This work was undertaken thanks in part to funding from NSERC, the Canada First Research Excellence Fund, the FRQNT, the U.S. Army Research Laboratory and the U.S. Army Research Office under contract/grant number W911NF-17-S-0008.
\end{acknowledgments}

\addcontentsline{toc}{section}{References}
\bibliography{biblio}{}

\appendix

\begin{figure*}[t]
	\centering
    \includegraphics[width=2\columnwidth]{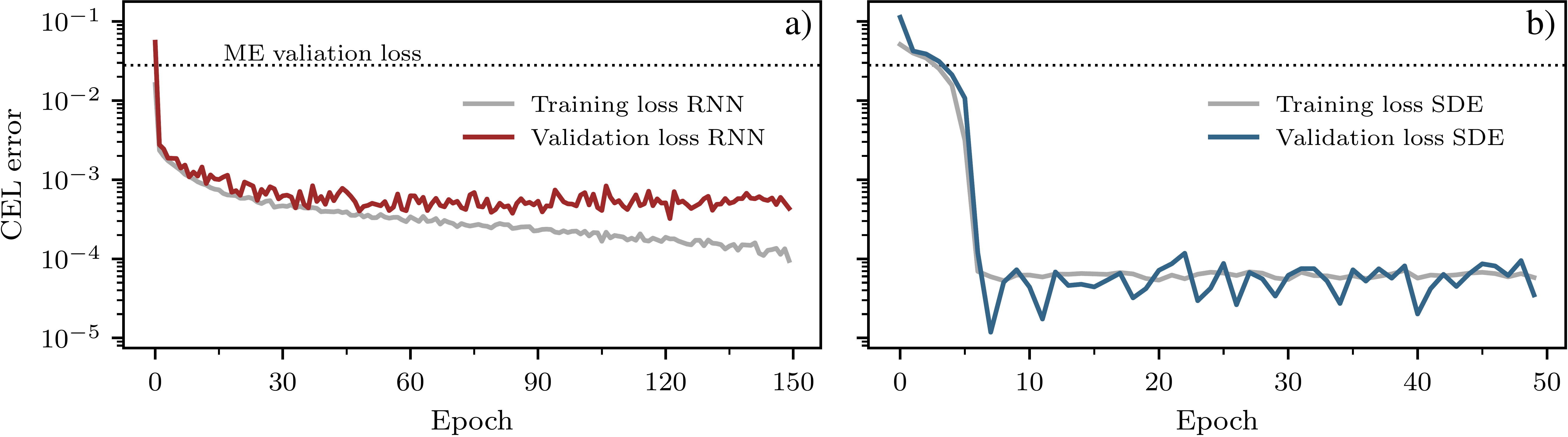}
	\caption{Training of our machine-learning models on numerically generated data.
	a) Cross entropy loss (CEL) error of a GRU with 16 hidden dimensions.
	The model overfits the training data containing 1.2 million weak-measurement time-series after more than about 120 epochs.
	b) Stochastic differential equation (SDE-learning) model in the form of~\cref{eq:SME} with free parameters $\{H_R,L,\eta\}$ trained on the same inputs as in a).
	This model has the same form as the equation used to generate the data and does not tend to significantly overfit the training data, but rather reaches a plateau for the CEL and the associated model parameters.
	We note that unlike the validation loss error, the training loss error is extracted directly from the training process where the model is updated at every mini-batch. 
	The dotted line represents the validation cross entropy loss error of a naive prediction given by the master equation (ME) using the true $H_R$ and $L$ operators.
	Both models rapidly learn to beat this prediction by making use of the weak measurements.
	We note that these ML trainings are qualitatively similar when using experimental data acquired from our superconducting qubit setup.
    }
	\label{fig:MLtraining}
\end{figure*}

\section{Recurrent Neural Networks}\label{appendix:NN}
A recurrent neural network (RNN) is formed of a unit cell that is repeated at every new time step of the input data $\Delta M_t$, producing an output $\bm{h}_{t}$ known as a hidden state.
This hidden state acts as a memory of the previous inputs and is combined with the next input of the time-series $\Delta M_{t+\Delta t}$ to allow information earlier in the sequence to influence outputs further down in the sequence.
As such, the RNN implements a non-linear function, defined in terms of the network parameters $\bm{W}$, that maps the previous hidden state and current input to the next hidden state.
Importantly, the function implemented by the RNN is fully differentiable with respect to these parameters such that they can be updated at each training iteration.
This process allows to bring the network output closer to the desired output.
It is done by taking the gradients of a chosen loss function $\mathcal{L}$ with respect to the network parameters using back-propagation and then updating them to minimize the loss function via gradient descent~\cite{Goodfellow-et-al-2016}
\begin{equation}
    \bm{W} \rightarrow \bm{W} - \zeta \frac{\partial\mathcal{L}}{\partial \bm{W}},
    \label{eq:backprop}
\end{equation}
where $\zeta$ is the learning rate.
In the main text, we are using a specific RNN architecture named gated recurrent unit (GRU).
After exploration of different well-known recurrent architecture that avoid the vanishing and exploding gradients problem, the GRU was found to be the most effective for the task at hand.

The specific function implemented by the GRU can be separated into three gates that are used to decide what information should be passed to the output.
They are named reset $r_t$, update $z_t$ and new $n_t$ gates and are expressed as~\cite{cho_learning_2014}
\begin{equation}\label{eq:GRUgates}
	\begin{split}
		r_t &= \sigma\left(W^{M,r}\Delta M_t+b^{M,r}+W^{h,r}h_{t-\Delta t}+b^{h,r}\right),\\
		z_t &= \sigma\left(W^{M,z}\Delta M_t+b^{M,z}+W^{h,z}h_{t-\Delta t}+b^{h,z}\right),\\
		n_t &= \tanh{\left(W^{M,n}\Delta M_t+b^{M,n}+r_t\ast\left(W^{h,n}h_{t-\Delta t}+b^{h,n} \right)\right)},\\
	\end{split}
\end{equation}
where $\bm{W}$ and $\bm{b}$ are the free weights and biases parameters optimized during the training stage,
$\sigma(\cdot)$ is the sigmoid function and $\ast$ is the Hadamard product.
The hidden state is then computed by combining these three gates in the following way
\begin{equation}
    h_t = \left(1-z_t\right)\ast n_t + z_t \ast h_{t-\Delta t}.
\end{equation}
Further information about the origin, possible performances and limitations of this architecture can be found in~\cite{cho_learning_2014, Goodfellow-et-al-2016, lipton_critical_2015}.

\section{Machine-learning trainings}\label{appendix:MLtrainings}
In~\cref{fig:MLtraining}, we present a typical training of a GRU and a SDE-learning model on numerically generated data.
The quantum trajectories and their associated weak-measurement data $\bm{\Delta M}_{t}$ were obtained numerically by integrating the stochastic master equation of~\cref{eq:SME} for a total time $T=\SI{8}{\micro s}$ for different noise realizations.

\Cref{fig:MLtraining}a) presents the training of the neural network illustrated in~\cref{fig:Overview} with a GRU containing 16 hidden dimensions and implemented in PyTorch~\cite{NEURIPS2019_9015}.
We use the loss function of~\cref{eq:LossFunction4L} with weights $w_\mathrm{posit}=0.36,\, w_\mathrm{prep}=1.7,\, w_\mathrm{\Delta M}=2.1$.
The training is done using the Adam optimizer~\cite{kingma_adam_2017} with a learning rate $\zeta=0.001$ and in batches of 1024 trajectories.
We observed that the training quality does not strongly depend on the specific values of these hyperparameters.
We see that, within the first few epochs, the GRU learns to do better than the average quantum trajectory prediction, which is given by integrating the master equation (ME) with the true parameters.
The ME is given by the first line of the SME~\cref{eq:SME}, i.e. by dropping the stochastic part.
The high expressive power of the GRU allows it to overfit the training data after enough training epochs, which is seen in~\cref{fig:MLtraining}a) by the increase of the validation loss.

\Cref{fig:MLtraining}b) presents the training of a stochastic differential equation integrator (SDE-learning) model that we implemented in PyTorch using the Milstein scheme presented in~\cite{kloeden_numerical_1994}.
The specific SDE used for our model is~\cref{eq:SME}, which is the same as the equation integrated to generate our weak-measurement data, thus allowing the SDE-learning to learn the quantum dynamics with high accuracy.
The small number of parameters of the SDE model allows us to train an ensemble of these models in parallel that each have their own, randomly initialized, free parameters.
For example here, 100 models are trained simultaneously on a single Nvidia GeForce 2080Ti with 11GB of memory.
\Cref{fig:MLtraining}b) shows the best model out of this ensemble.
The SDE model is also optimized using Adam and the same values for the learning rate and batch size.
Since the SDE model automatically satisfies the quantum-mechanical properties we were trying to encourage with additional loss-function terms, the SDE-learning is trained solely on the cross-entropy loss~\cref{eq:LossFunctionCEL}.
We highlight the fact that the SDE-learning model is given the same inputs, which are realistically available in an experiment, as the GRU.
These inputs are the weak-measurement time-series tensors $\bm{\Delta M}_{t}$, together with their associated preparation $\vec{p}$ and final projective measurement $\vec{m}$ in the form of one-hot encoded vectors, as illustrated in~\cref{fig:Overview}.

\section{Limitations of the finite sampling time}\label{appendix:CoarseGraining}

\begin{figure}[t]
	\centering
    \includegraphics[width=\columnwidth]{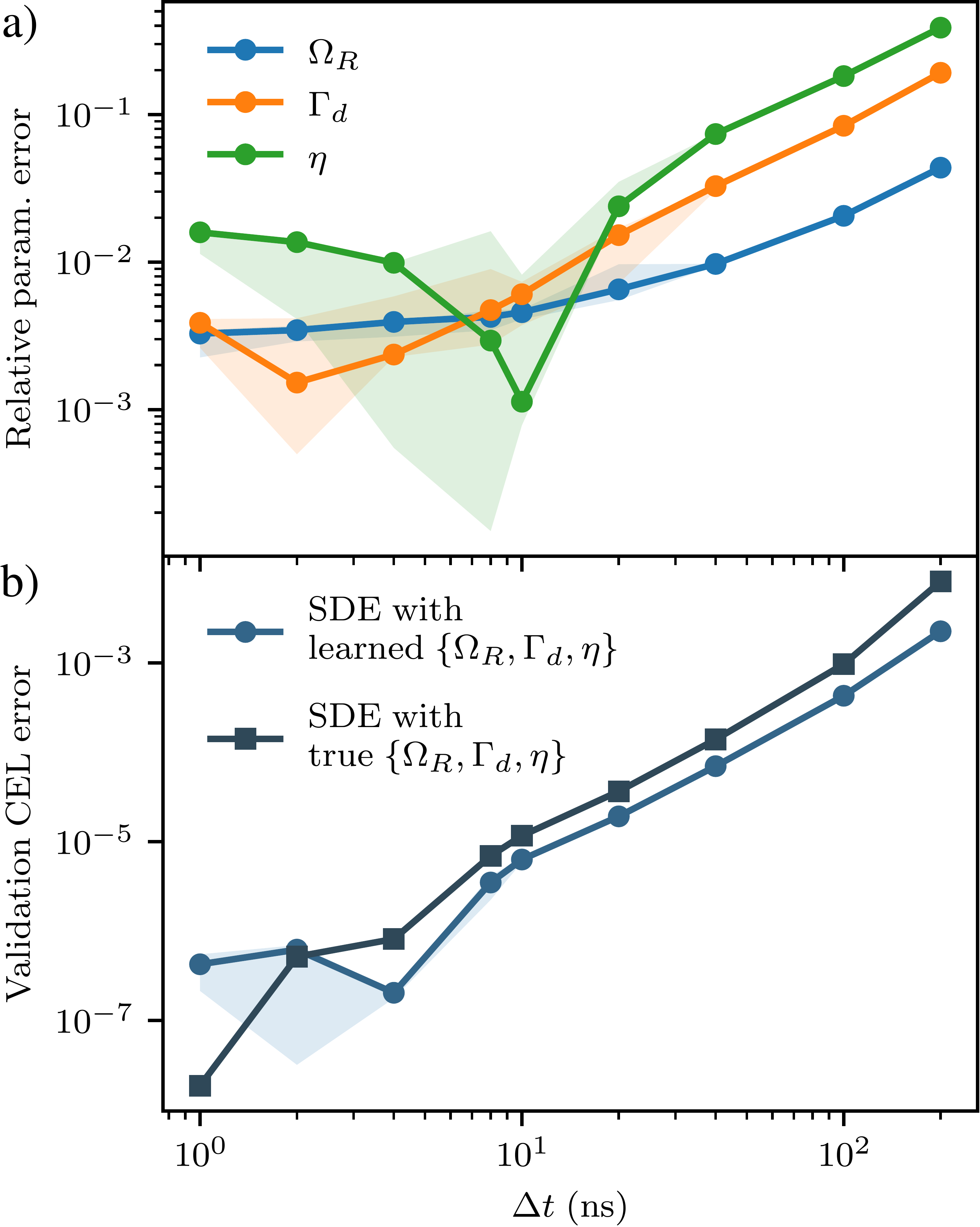}
	\caption{
	Characterizing quantum dynamics with coarsed grained trajectories.
	a) The error of the learned parameters grows with the size of the weak-measurement time step $\Delta t$.  
	This implies that $\Delta t$ imposes a limitation on the accuracy at which the SDE model can learn true device parameters.
	b) The cross entropy loss (CEL) error on the validation grows proportionally to $\Delta t$, as caused by numerical integration errors of the Milstein scheme.
	All the curves with circles represent the 25\% quantile of the distribution of trained SDE models on the data, with the shaded regions limited by the 0\% and 50\% quantiles.
	The true parameters are chosen to be the ones obtained from the calibration of the superconducting qubit of the main text, namely
	$\Omega_R = \SI{1.395}{\micro s^{-1}}$, $\Gamma_d = \SI{1.176}{\micro s^{-1}}$ and $\eta=0.1469$.
	}
	\label{fig:CoarseGrainingParams}
\end{figure}

The finite step size $\Delta t$ of the time-series, experimentally limited by the sampling time of the weak measurements, intrinsically limits our ability to reconstruct quantum trajectories by integrating a stochastic master equation.
Indeed, some information is lost by integrating the weak-measurement signal in time steps $\Delta t$.
Throughout this work, we use $\Delta t=\SI{0.04}{\micro s}$ for the experimental and numerical data.
This coarse-graining of the gathered $\Delta t=\SI{0.001}{\micro s}$ time-series is motivated by two practical considerations.
First, the computing time and memory requirements associated with having a small time step are important for long sequences. 
Using this time step allows us to train all of our machine-learning models using less than a couple hours and 11~GB of GPU RAM.
Second, we want to avoid the correlations in the time-series that are due to the finite memory time of the readout cavity $(1/\kappa\approx\SI{0.10}{\micro s})$, since our SDE models assume uncorrelated weak measurements. 

In order to explore the effect of this coarse graining on the ability of the SDE-learning model to learn accurate device parameters, we train this model on synthetic data coarse grained to different $\Delta t$.
The training data contains 75~000 trajectories obtained by integrating the following SME
\begin{align}
\begin{split}
    \mathrm{d}\rho_t = & -i\frac{\Omega_R}{2} \left[ \sigma_x, \rho_t\right] \mathrm{d}t + \frac{\Gamma_d}{2}\mathcal{D}[\sigma_z]\rho_t \mathrm{d}t \\
    & + \sqrt{\frac{\eta\Gamma_d}{4}}\left(\mathcal{H}[\sigma_z]\rho_t \mathrm{d}W_t^\mathrm{I} +  \mathcal{H}[-i\sigma_z]\rho_t \mathrm{d}W_t^\mathrm{Q}\right),
    \label{eq:SMErelax}
\end{split}
\end{align}
for variable durations $T$ between $\SI{0}{\micro s}$ and $\SI{8}{\micro s}$ and with a resolution $\Delta t=\SI{0.001}{\micro s}$.
From these trajectories, the other training datasets are generated by combining weak measurements in order to coarse grain the data to larger time steps, up to $\Delta t=\SI{0.2}{\micro s}$.
We do the same for the validation set with half as many trajectories as the training set.
We note that we generate all of these trajectories with a Milstein integration scheme, which corresponds to the scheme used by our SDE-learning model.

In~\cref{fig:CoarseGrainingParams}, we quantify the effect of the integration errors on the SDE-learning achievable accuracy for different coarse graining of the weak-measurement time-series.
We present both the relative error of the learned parameters $\{\Omega_R, \Gamma_d, \eta\}$ and the associated prediction accuracy of the SDE model, as given by the validation cross-entropy loss (CEL) error.
\Cref{fig:CoarseGrainingParams}a) shows that we can learn the true physical rates and quantum measurement efficiency within a relative error close to 1\% or less.
We attribute this finite precision to the limited size of the training dataset and the use of mini-batches during training, which both introduce statistical sampling errors.
However, this precision gets worse with increasing $\Delta t$, thus confirming that the integration errors introduced by the coarse-graining is a limitation for the SDE-learning model, and more generally to any physical modeling of the data involving the numerical integration of a differential equation. 

In~\cref{fig:CoarseGrainingParams}b), we present the isolated effect of the integration errors on the cross entropy loss achieved by a SDE model.
To do so, we integrate the validation trajectories using the true values of the parameters $\{\Omega_R, \Gamma_d, \eta\}$ for the differently coarse-grained datasets and compare these trajectories with the expected projective measurement outcomes.
We see that the CEL error grows proportionally to the size of $\Delta t$ for the SDE model with true parameters.
When leaving these same parameters $\{\Omega_R, \Gamma_d, \eta\}$ free for the SDE model to learn during training, we observe that the SDE-learning model is able to perform better at predicting the measurement outcomes of the validation set.
We attribute this effect to the fact that the model has the freedom to find parameters that account for part of the information-loss due to coarse-graining the data.
For example, learning a slightly lower quantum measurement efficiency $\eta$ than the true value, on coarse-grained data, allows the SDE model to achieve a better CEL. 
Consequently, the finite size of $\Delta t$ can bias the parameter estimation and thus limit the ability of the SDE-learning to extract accurate device parameters.
As such, we use this bias found in training on numerical data as a lower bound on the error associated with learning the device parameters on experimental data with $\Delta t=\SI{0.04}{\micro s}$.
We use this error for the error bars of the SDE-learning in~\cref{fig:Params} of the main text, since it is larger than the standard deviation of the parameters learned in the ensemble of 100 SDE models.

\section{Parameter values}\label{appendix:Params}

For completeness, we present here the numerical values illustrated in Fig.~\ref{fig:Params} of the main text.

\begin{table}[h]
\begin{tabularx}{1\columnwidth}{c c c c c c c}
\cline{1-7}
\noalign{\vskip\doublerulesep
         \vskip-\arrayrulewidth}
\cline{1-7}
                      &                                                                   & \begin{tabular}[c]{@{}c@{}}True\\ or\\ Calib.\end{tabular}  & \begin{tabular}[c]{@{}c@{}}SDE\\ $\{H_R,L,\eta\}$\end{tabular} & \begin{tabular}[c]{@{}c@{}}RNN\\ + SDE\\ $\{H_R,L,\eta\}$\end{tabular} & \begin{tabular}[c]{@{}c@{}}RNN\\ + SDE\\ $\{\Omega_R,\Gamma_d,\eta\}$\end{tabular} & \begin{tabular}[c]{@{}c@{}}RNN +\\ Binning \end{tabular} \\ \cline{1-7}
\multicolumn{1}{l|}{\parbox[b]{4mm}{\multirow{4}{*}{\rotatebox[origin=c]{90}{Numerical data}}}} & \begin{tabular}[c]{@{}c@{}}$\Omega_R/2\pi$\\ (MHz)\end{tabular}   & 0.222   & 0.220                                                       & 0.221                                                               & 0.221                                                                             & 0.215                                                   \\
\multicolumn{1}{l|}{}                      & \begin{tabular}[c]{@{}c@{}}$\Gamma_d / 2\pi$\\ (MHz)\end{tabular} & 0.187   & 0.194                                                       & 0.195                                                               & 0.193                                                                             & 0.200                                                   \\
\multicolumn{1}{l|}{}                      & \begin{tabular}[c]{@{}c@{}}$\eta$ \\ (\%)\end{tabular}            & 14.7   & 13.4                                                        & 13.5                                                                & 13.7                                                                              & 12.4                                                    \\ \cline{2-7} 
\multicolumn{1}{l|}{}                      & \begin{tabular}[c]{@{}c@{}}Cross\\ entropy\end{tabular}           & 0.64510 & 0.64512                                                      & 0.64513                                                              & 0.64511                                                                            & 0.64527                                                 \\
\cline{1-7}
\noalign{\vskip\doublerulesep
         \vskip-\arrayrulewidth}
\cline{1-7}
\multicolumn{1}{l|}{\parbox[b]{4mm}{\multirow{4}{*}{\rotatebox[origin=c]{90}{Experimental data}}}} & \begin{tabular}[c]{@{}c@{}}$\Omega_R/2\pi$\\ (MHz)\end{tabular}   & 0.222   & 0.228                                                        & 0.228                                                               & 0.232                                                                             & 0.213                                                   \\
\multicolumn{1}{l|}{}                      & \begin{tabular}[c]{@{}c@{}}$\Gamma_d / 2\pi$\\ (MHz)\end{tabular} & 0.187   & 0.182                                                        & 0.176                                                               & 0.184                                                                             & 0.173                                                   \\
\multicolumn{1}{l|}{}                      & \begin{tabular}[c]{@{}c@{}}$\eta$\\ (\%)\end{tabular}             & 14.7   & 14.5                                                        & 13.8                                                                & 13.3                                                                              & 20.2                                                    \\ \cline{2-7} 
\multicolumn{1}{l|}{}                      & \begin{tabular}[c]{@{}c@{}}Cross\\ entropy\end{tabular}           & --      & 0.6478                                                       & 0.6489                                                               & 0.6504                                                                             & 0.6517                                                  \\
\cline{1-7}
\noalign{\vskip\doublerulesep
         \vskip-\arrayrulewidth}
\cline{1-7}
\end{tabularx}
\label{tab:ResultsExpGen}
\caption{Different approaches to learning device parameters from weak measurements in the cases of synthetic and experimental data.
    The RNN of the last three columns is a single model trained on the weak measurement data and the parameters are extracted from its output quantum trajectories.
    The cross entropy is a measure of the accuracy of the model and associated parameters to describe the observed measurement outcomes.
    Note that the binning approach assumes the same physical model as the $\mathrm{RNN+SDE\;\{\Omega_R,\Gamma_d,\eta\}}$ model, i.e. it assumes that $H_R=\Omega_R/2 \sigma_x$ and $L=\sqrt{\Gamma_d/2}\sigma_z$ in~\cref{eq:SME}.
    }
\end{table}

\end{document}